# Cross-correlation properties of cyclotomic sequences

Kai Cai[1,2]*, Rongquan Feng[1,3], and Zhiming Zheng[1]

1. LMAM, School of Mathematical Sciences, Peking University, Beijing, China, 100871

2. Dept. of Electronic Engineering Tsinghua University, Beijing, China, 100084

3. State Key Laboratory of Information Security
(Graduate School of Chinese Academy of Sciences), Beijing, China, 100039

E-mails: {caik@tsinghua., fengrq@math.pku.,zzheng@math.pku.}edu.cn

**Abstract**

Sequences with good correlation properties are widely used in engineering applications, especially in the area of communications. Among the known sequences, cyclotomic families have the optimal autocorrelation property. In this paper, we decide the cross-correlation function of the known cyclotomic sequences completely. Moreover, to get our results, the relations between the multiplier group and the decimations of the characteristic sequence are also established for an arbitrary difference set.

## 1 Introduction

Let $D$ be a subset of $\mathbb{Z}_N$, the *characteristic* sequence $s$ of $D$ is a binary one which is defined by

$$s_i = \begin{cases} 1, & \text{if } i \,(\text{mod } N) \in D, \\ 0, & \text{otherwise,} \end{cases}$$

where $s_i$ is the $i$-th component of $s$. $D$ is called the *characteristic set* or *support* of $s$. Let $s = (s_0, s_1, \ldots, s_{N-1})$ and $t = (t_0, t_1, \ldots, t_{N-1})$ be two binary sequences of period $N$ with characteristic sets $B$ and $C$, respectively. The period *cross-correlation function* between $s$ and $t$ is defined as

$$C_{s,t}(w) = \sum_{i=0}^{N-1} (-1)^{s_{i+w}-t_i}$$

---

*Corresponding author

for $w = 0, 1, \ldots, N - 1$, where the operation in the subscript is done modulo $N$. From now on, all the operations related to the subscript will be done modulo a particular integer. $C_{s,t}(w)$ can be determined by the *difference function* of $B$ and $C$ defined as $d_{B,C}(w) = |(B+w) \cap C|$, where $B+w = \{x+w \,|\, x \in B\}$. When $|B| = |C|$, it can be got easily that

$$C_{s,t}(w) = N - 4(|B| - d_{B,C}(w)). \tag{1.1}$$

When $B = C$ and then $s = t$, we denote $d_{B,C}(w)$ and $C_{s,t}(w)$ by $d_B(w)$ and $C_s(w)$, respectively, with the latter being known as the *autocorrelation function* of $s$. The above equation (1.1) then becomes $C_s(w) = N - 4(|B| - d_B(w))$ (see [4]).

Periodic binary sequences with good correlation properties have important applications in various areas of engineering, such as data acquisition and analysis system [10], detection system [11, 21], cryptology [6], and especially communications [5]. Many applications need binary sequences that have good autocorrelation properties, while in multiple-terminal system identification and code-division multiple-access (CDMA) system, there need sequences with both good autocorrelation and cross-correlation properties.

Let $GF(q)$ be the finite field with $q$ elements, and let $d$ divide $q - 1$, i.e., $q = df + 1$ for some integer $f$. Let $\alpha$ be a primitive element of $GF(q)$, define $C_0^{(d,q)} = \langle \alpha^d \rangle = \{1, \alpha^d, \alpha^{2d}, \ldots, \alpha^{(f-1)d}\}$ and $C_i^{(d,q)} = \alpha^i C_0^{(d,q)}$, for $i = 0, 1, \ldots, d-1$. These $C_i^{(d,q)}$'s are called *cyclotomic classes* of order $d$. The *cyclotomic numbers* of order $d$ with respect to $GF(q)$ are defined as $(i, j) = |(C_i^{(d,q)} + 1) \cap C_j^{(d,q)}|$. The cyclotomic numbers [17] which will be used in this paper are given as follows.

(1) If $q = 2f + 1$ with even $f$, then

$$(0, 0) = (f - 2)/2, \text{ and } (0, 1) = (1, 0) = (1, 1) = f/2. \tag{1.2}$$

(2) If $q = 2f + 1$ with odd $f$, then

$$(0, 1) = (f + 1)/2, \text{ and } (0, 0) = (1, 0) = (1, 1) = (f - 1)/2. \tag{1.3}$$

(3) If $q = 6f + 1 = x^2 + 27$, then see Table 1. In this table, the numbers at the left column and at the top row indicate the indexes of rows and columns respectively. The cyclotomic number $(i, j)$ is the entry at the row with index $i$ and at the column with index $j$, where $A = (q - 11 - 8x)/36$, $B = (q + 37 - 2x)/36$, $C = (q + 1 + 16x)/36$, $D = (q - 35 - 2x)/36$, $E = (q + 13 + 4x)/36$, $F = (q - 23 + 4x)/36$ and $G = (q + 1 - 2x)/36$.

**Table 1:** The cyclotomic numbers for $q = 6f + 1 = x^2 + 27$.

| $i\backslash j$ | 0 | 1 | 2 | 3 | 4 | 5 |
|---|---|---|---|---|---|---|
| 0 | A | B | B | C | D | D |
| 1 | E | F | G | D | B | G |
| 2 | F | G | E | D | G | B |
| 3 | A | E | F | A | E | F |
| 4 | E | D | G | B | F | G |
| 5 | F | G | D | B | G | E |

Let $G$ be an abelian group of order $v$. A $(v, k, \lambda)$-*difference set* in $G$ is a $k$-subset $D \subseteq G$ such that the equation $x - y = g$ has just $\lambda$ solutions $(x, y) \in D \times D$ for each nonzero element $g$ of $G$. A difference set in the cyclic group $\mathbb{Z}_v$ is also called *cyclic*. Cyclotomic classes can be used to construct difference sets [1, 17]. Those difference sets constructed by cyclotomic classes are called *cyclotomic* difference sets. Moreover, the sequences constructed by cyclotomic difference sets are called *cyclotomic* sequences.

Many constructions of binary sequences with good correlation properties were given for the propose of application. m-sequences have optimal autocorrelation values, but the cross-correlations between some pairs may not fit for some particular usages. Thus many authors took efforts to seek the preferred pairs of m-sequences [2, 8]. For more recent results on the cross-correlations of m-sequences, the reader is referred to Dobbertin *et. al* [3] and Ness *et. al* [12]. Meanwhile, many sequence families having both low nontrivial auto and cross-correlation values were also proposed, including them, bent and Gold sequences, as well as the small and large families of Kasami sequences are the famous ones [13, 15]. For the new constructions, the reader can see [20] and its references. In this paper, we concentrate on the cross-correlation of the known family of cyclotomic sequences that have perfect autocorrelations.

The following of the paper is organized as follows. Relations between multipliers of a difference set and the decimations of its characteristic sequence are established in Section 2. The cross-correlation functions of the known cyclotomic sequences are decided in Section 3. Concluding remarks are in Section 4.

## 2 Multiplier and decimation

Let $D$ be a $(v, k, \lambda)$-difference set in an abelian group $G$ and let $r$ be a positive integer relatively prime to $v$. If $rD = D + g$ for some $g \in G$, then we call $r$ a (numerical) *multiplier* of $D$, where $rD = \{rd \,|\, d \in D\}$ and $D + g = \{d + g \,|\, d \in D\}$. All multipliers of $D$ form a subgroup $M$ of $\text{Aut}(G)$, which is called the *multiplier group* of $D$, where $\text{Aut}(G)$ is the automorphism group of the group $G$.

Let $s = (s_0, s_1, \ldots, s_{N-1})$ be a binary sequence of period $N$, and $r$ an integer relatively prime to $N$. Then $s[r] = (s_0, s_r, s_{2r}, \ldots, s_{(N-1)r})$ is also a binary sequence of period $N$. We call this sequence the *r-decimation* of $s$. Let $T$ be the left cyclic shift transformation on sequences. That is, for any sequence $s = (s_0, s_1, \ldots, s_{N-1})$, $Ts = (s_1, \ldots, s_{N-1}, s_0)$. Define $T^i s = T(T^{i-1}s)$ for $2 \le i \le N-1$. All these sequences $T^i s$'s, $0 \le i \le N-1$, are called *phases* of $s$, where $T^0 s = s$ and $T^1 s = Ts$. Throughout this paper we consider a sequence and its phases are the same one. The following proposition is obvious from the definitions.

**Proposition 2.1** *Let $D$ be a difference set in $\mathbb{Z}_N$ with characteristic sequence $s$, and $r$ an integer relatively prime to $N$. Then $s[r]$ is a phase of $s$ if and only if $r$ is a multiplier of $D$.*

Two sequences are said to have the same autocorrelation spectrum if their autocorrelation functions have the same image and each value in the image occurs the same times.

**Proposition 2.2** *For any $N$-period sequence $s$ and an integer $r$ relatively prime to $N$, the two sequences $s$ and $s[r]$ have the same autocorrelation spectrum.*

**Proof** The result is immediate from the fact that

$$C_{s[r]}(w) = \sum_{i=0}^{N-1}(-1)^{s_{r(i+w)}-s_{ri}} = \sum_{\ell=0}^{N-1}(-1)^{s_{\ell+rw}-s_\ell} = C_s(rw)$$

and that $\varphi : w \mapsto rw$ is a permutation on $\mathbb{Z}_N$. □

It is well-known that each automorphism of the cyclic group $\mathbb{Z}_N$ is of the form $z \mapsto rz$ for any $z \in \mathbb{Z}_N$, where $r$ is an integer relatively prime to $N$. We denote such an automorphism by $r$ again. For the sake of convenience, we abuse notations by using the same letter $r$ for an integer which is relatively prime to $N$ and for an automorphism of the group $\mathbb{Z}_N$ derived from $r$.

**Lemma 2.3** *Let $D$ be a difference set in $\mathbb{Z}_N$ with characteristic sequence $s$ and let $M$ be the multiplier group of $D$. Suppose that*

$$\mathrm{Aut}(\mathbb{Z}_N) = \cup_{i=0}^{n-1} r_i M$$

*is the coset decomposition of $\mathrm{Aut}(\mathbb{Z}_N)$ with respect to $M$, where $r_0 = 1$ and $n = \phi(N)/|M|$. Here $\phi$ is the Euler $\phi$-function. Then*
 (1) *$s, s[r_1], \ldots, s[r_{n-1}]$ are all the decimations of $s$.*
 (2) *The characteristic set of $s[r_i]$ is $r_i^{-1}D$, for $i = 0, 1, \ldots, n-1$.*

**Proof** (1) For any $r$ which is relatively prime to $N$, let $r = r_i m$ for some $r_i$ and some $m \in M$. Then by Proposition 2.1 there exist $x, y \in \mathbb{Z}_N$ such that $s[r] = s[r_i m] = s[r_i][m] = s[m][r_i] = (T^x s)[r_i] = T^y(s[r_i])$. So any decimation of $s$ is a phase of $s[r_i]$. On the other hand, if $s[r_i]$ is a phase of $s[r_j]$ with $i \neq j$, then $s[r_i r_j^{-1}]$ is a phase of $s$. Thus $r_i r_j^{-1} \in M$, that is, $r_i$ and $r_j$ are in the same coset, which is a contradiction. Hence $s[r_i]$ and $s[r_j]$ are different sequences.

(2) Let $u = (u_0, u_1, \ldots, u_{N-1})$ be the characteristic sequence of $r_i^{-1} D$. Then $u_k = 1 \Leftrightarrow k \in r_i^{-1} D \Leftrightarrow r_i k \in D$. Thus $u = s[r_i]$. □

Those sets $r_i^{-1} D$ in Lemma 2.3 are called the *conjugates* of $D$. An $m$-sequence of degree $n$ is the characteristic sequence of a Singer $(2^n - 1, 2^{n-1} - 1, 2^{n-2} - 1)$-difference set. The multiplier group of this Singer difference set has order $n$ (see [14]). By Lemma 2.3, one can get easily that the number of $m$-sequences of degree $n$ is $\phi(2^n - 1)/n$.

**Remark 2.4** *For any subset of $\mathbb{Z}_N$, not necessarily a difference set, its multipliers can be defined by the same manner and the above results also hold.*

A binary sequence $s$ of period $N$ is called a *perfect* one if $C_s(w) = -1$ for all $w \neq 0$. The well-known families of perfect sequences mainly include $m$-sequences and the characteristic sequences of Paley difference sets, twin prime difference sets and Hall difference sets ([7]-[14]), which are called *Paley sequences*, *twin prime sequences* and *Hall sequences*, respectively. The definitions of these difference sets and their multiplier groups ([1], [17]) are listed below.

(1) Paley difference sets: Let $q \equiv 3 \pmod{4}$ be a prime and $G = (GF(q), +)$. $D = C_0^{(2,q)}$ and $M = D$.

(2) Twin prime difference sets: Let both $q$ and $q + 2$ be primes and $G = (GF(q) \times GF(q+2), +)$. $D = (C_0^{(2,q)} \times C_0^{(2,q+2)}) \cup (C_1^{(2,q)} \times C_1^{(2,q+2)}) \cup (GF(q) \times \{0\})$ and $M = (C_0^{(2,q)} \times C_0^{(2,q+2)}) \cup (C_1^{(2,q)} \times C_1^{(2,q+2)})$.

(3) Hall difference sets: Let $q$ be a prime and $q = x^2 + 27 \equiv 1 \pmod{6}$. $G = (GF(q), +)$. $D = C_0^{(6,q)} \cup C_1^{(6,q)} \cup C_3^{(6,q)}$ and $M = C_0^{(6,q)}$.

**Remark 2.5** *By the Chinese Remainder Theorem, $\mathbb{Z}_{q(q+2)} \simeq GF(q) \times GF(q+2)$ under the isomorphism $x \mapsto (x \pmod{q}, x \pmod{(q+2)})$. So all the above three difference sets are cyclic.*

## 3  Cross-correlation properties of cyclotomic sequences

Among the known sequences families, perfect sequences have the best autocorrelation properties. But for applications, it requires sequences with both good autocorrelation properties and good cross-correlation properties. Unfortunately, such requirement seems not easy to be satisfied by existing of restricting relations among the peak cross-correlation magnitude, the peak out-of-phase autocorrelation magnitude,and the period and the number of sequences in a sequence set [15]. Many efforts of investigating the cross-correlation function of $m$-sequences have been taken [16, 18, 19]. Here, we study the cross-correlation properties of the other three families of perfect sequences mentioned in Section 2. It shows that the cross-correlation function of such sequences can be completely decided and all the sequence pairs of the families have lower cross-correlation values.

**Theorem 3.1** *Let $q \equiv 3 \,(\mathrm{mod}\, 4)$ be a prime, and $D$ the Paley difference set in $GF(q)$ with the characteristic sequence $s$. Then*

(1) *$s$ has two decimations, say $s$ and $t$.*

(2) *Both $s$ and $t$ are perfect sequences.*

(3) *$C_{s,t}(w)$ is 3-valued. More precisely, $C_{s,t}(0) = 2 - q$ and $C_{s,t}(w)$ is $-1$ or $3$ when $w \neq 0$.*

**Proof** (1) By Lemma 2.3, $s$ has $\frac{\phi(q)}{|M|} = \frac{q-1}{(q-1)/2} = 2$ decimations.

(2) It is clear by Proposition 2.2.

(3) The conjugates of $D$ are $D = C_0^{(2,q)}$ and $\widetilde{D} = C_1^{(2,q)}$. Also it is clear that $d_{D,\widetilde{D}}(0) = 0$. When $w \neq 0$,

$$\begin{aligned}
d_{D,\widetilde{D}}(w) &= |(C_0^{(2,q)} + w) \cap C_1^{(2,q)}| \\
&= |(w^{-1} C_0^{(2,q)} + 1) \cap w^{-1} C_1^{(2,q)}| \\
&= \begin{cases} (0,1), & \text{if } w^{-1} \in C_0^{(2,q)}, \\ (1,0), & \text{if } w^{-1} \in C_1^{(2,q)}. \end{cases}
\end{aligned}$$

Noticing that $(C_i^{(2,q)})^{-1} = C_i^{(2,q)}, i = 0, 1$, and $q = 2f + 1$ with odd $f$, we have by Equation (1.3) that

$$C_{s,t}(w) = q - 4\left(\frac{q-1}{2} - d_{D,\widetilde{D}}(w)\right) = \begin{cases} 2 - q, & \text{if } w = 0, \\ 3, & \text{if } w \in C_0^{(2,q)}, \\ -1, & \text{if } w \in C_1^{(2,q)}. \end{cases}$$

□

**Example 3.2** *The Paley sequence of period 19 is*

$$s = (0, 1, 0, 0, 1, 1, 1, 1, 0, 1, 0, 1, 0, 0, 0, 0, 1, 1, 0)$$

*and its 2-decimation is*

$$t = (0, 0, 1, 1, 0, 0, 0, 0, 1, 0, 1, 0, 1, 1, 1, 1, 0, 0, 1).$$

We have that $C_{s,t}(0) = -17$, $C_{s,t}(w) = -1$ when $w \in \{2, 3, 8, 10, 12, 13, 14, 15, 18\}$ and $C_{s,t}(w) = 3$ for other $w$'s.

**Theorem 3.3** *Let both $q$ and $q+2$ be primes, $D$ the twin prime difference set in $GF(q) \times GF(q+2)$ and let $s$ be the characteristic sequence of $D$. Then*
  *(1) $s$ has two decimations, say $s$ and $t$.*
  *(2) Both $s$ and $t$ are perfect sequences.*
  *(3) When $q = 3$, $C_{s,t}(w)$ is 4-valued, and when $q \neq 3$, $C_{s,t}(w)$ is 5-valued.*

**Proof** (1) and (2) are similar as in Theorem 3.1. We prove only (3) now. For simplicity, denote $C_i^{(2,q)}$ and $C_i^{(2,q+2)}$ by $C_i$ and $C_i'$, respectively, $i = 0, 1$. The conjugates of $D$ are $D = (C_0 \times C_0') \cup (C_1 \times C_1') \cup (GF(q) \times \{0\})$ and $\widetilde{D} = (C_0 \times C_1') \cup (C_1 \times C_0') \cup (GF(q) \times \{0\})$. We distinguish the following 4 cases to compute the difference function $d_{D,\widetilde{D}}(w)$.

1) If $w = (0,0)$, then $d_{D,\widetilde{D}}(w) = q$.

2) If $w = (a, 0)$ with $a \neq 0$, then
$$d_{D,\widetilde{D}}(w) = |(D + (a,0)) \cap \widetilde{D}|$$
$$= |C_0'| \cdot |(C_0 + a) \cap C_1| + |C_1'| \cdot |(C_1 + a) \cap C_0| + q$$
$$= ((q^2 - 1)/4) + q.$$

3) If $w = (0, b)$ with $b \neq 0$, then
$$d_{D,\widetilde{D}}(w) = |(D + (0,b)) \cap \widetilde{D}|$$
$$= |C_0| \cdot |((C_0' + b) \cup \{b\}) \cap (C_1' \cup \{0\})|$$
$$+ |C_1| \cdot |((C_1' + b) \cup \{b\}) \cap (C_0' \cup \{0\})|.$$

Noticing that when $q \equiv 3 \pmod{4}$, $b \in C_0'$ implies $-b \in C_0'$ and $b \in C_1'$ implies $-b \in C_1'$ and that when $q \equiv 1 \pmod{4}$, $b \in C_0'$ implies $-b \in C_1'$ and $b \in C_1'$ implies $-b \in C_0'$, we have
$$d_{D,\widetilde{D}}(w) = ((q^2 - 1)/4) + q - 1.$$

4) If $w = (a, b) \neq (0, 0)$, then
$$d_{D,\widetilde{D}}(w) = |(D + (a,b)) \cap \widetilde{D}| = |(w^{-1}D + (1,1)) \cap w^{-1}\widetilde{D}|.$$

Let $w^{-1} \in C_i \times C_j'$, $(i, j = 0, 1)$. We have that
$$d_{D,\widetilde{D}}(w) = |((C_i + 1) \times (C_j' + 1)) \cap (C_i \times C_{1+j}')|$$
$$+ |((C_i + 1) \times (C_j' + 1)) \cap (C_{1+i} \times C_j')|$$
$$+ |((C_{1+i} + 1) \times (C_{1+j}' + 1)) \cap (C_i \times C_{1+j}')|$$
$$+ |((C_{1+i} + 1) \times (C_{1+j}' + 1)) \cap (C_{1+i} \times C_j')|$$
$$+ q - 1.$$

By straightforward calculations, whether $q \equiv 1 \pmod 4$ or $q \equiv 3 \pmod 4$, we always have

$$d_{D,\tilde{D}}(w) = \begin{cases} ((q+1)^2/4) - 2, & \text{if } w \in (C_0 \times C_0') \cup (C_1 \times C_1'), \\ ((q-1)^2/4) + q - 1, & \text{if } w \in (C_1 \times C_0') \cup (C_0 \times C_1'). \end{cases}$$

Thus, by Equation (1.1) and the fact that $(C_i)^{-1} = C_i$, $(C_i')^{-1} = C_i'$, $i = 1, 2$, we have

$$C_{s,t}(w) = \begin{cases} -q^2 + 2q + 2, & \text{if } w = (0,0), \\ 2q + 1, & \text{if } w = (a,0), a \neq 0, \\ 2q - 3, & \text{if } w = (0,b), b \neq 0, \\ -5, & \text{if } w \in (C_0 \times C_0') \cup (C_1 \times C_1'), \\ -1, & \text{if } w \in (C_1 \times C_0') \cup (C_0 \times C_1'). \end{cases}$$

Obviously, when $q \neq 3$, the five values of $C_{s,t}(w)$ can not coincide, which proves the theorem. $\square$

**Example 3.4** *The twin prime sequence of period 15 is*

$$s = (1,1,1,0,1,1,0,0,1,0,1,0,0,0,0)$$

*and its 7-decimation is*

$$t = (1,0,0,0,0,1,0,1,0,0,1,1,0,1,1).$$

*The cross-correlation function between $s$ and $t$ is given below:*

| $w$ | 0 | 1 | 2 | 3 | 4 | 5 | 6 | 7 | 8 | 9 | 10 | 11 | 12 | 13 | 14 |
|---|---|---|---|---|---|---|---|---|---|---|---|---|---|---|---|
| $C_{s,t}(w)$ | -1 | -5 | -5 | 3 | -5 | 7 | 3 | -1 | -5 | 3 | 7 | -1 | 3 | -1 | -1 |

**Theorem 3.5** *Let $q \equiv 1 \pmod 6$ be a prime and assume that $q$ can be decomposed as $q = x^2 + 27$. Let $D$ be the Hall difference set in $GF(q)$ and $s$ the characteristic sequence of $D$. Then*

*(1) $s$ has 6 decimations, say $s_0 = s, s_1, s_2, s_3, s_4, s_5$.*
*(2) All these $s_i$'s, $0 \leq i \leq 5$, are perfect sequences.*
*(3) $C_{s_i,s_j}(w)$ is 6-valued or 7-valued if $q \neq 31$.*

**Proof** We prove only the statement (3). Obviously, the conjugates of $D$ are $D_i = C_i^{(6,q)} \cup C_{1+i}^{(6,q)} \cup C_{3+i}^{(6,q)}$, $i = 0, 1, 2, 3, 4, 5$, where $D_0 = D$. For the difference function between $D_i$ and $D_j$, we have

$$d_{D_i,D_j}(w) = |(D_i + w) \cap D_j| = \begin{cases} |D_i \cap D_j|, & \text{if } w = 0, \\ |(w^{-1}D_i + 1) \cap w^{-1}D_j|, & \text{if } w \neq 0. \end{cases}$$

When $w \neq 0$, let $w^{-1} \in C_\ell^{(6,q)}$, $0 \leq \ell \leq 5$, then

$$\begin{aligned}
d_{D_i,D_j}(w) &= |(D_{i+\ell} + 1) \cap D_{j+\ell}| \\
&= |((C_{i+\ell}^{(6,q)} \cup C_{1+i+\ell}^{(6,q)} \cup C_{3+i+\ell}^{(6,q)}) + 1) \cap (C_{j+\ell}^{(6,q)} \cup C_{1+j+\ell}^{(6,q)} \cup C_{3+j+\ell}^{(6,q)})| \\
&= (i+\ell, j+\ell) + (i+\ell, 1+j+\ell) + (i+\ell, 3+j+\ell) \\
&\quad + (1+i+\ell, j+\ell) + (1+i+\ell, 1+j+\ell) \\
&\quad + (1+i+\ell, 3+j+\ell) + (3+i+\ell, j+\ell) \\
&\quad + (3+i+\ell, 1+j+\ell) + (3+i+\ell, 3+j+\ell).
\end{aligned}$$

Noticing that $(C_1^{(6,q)})^{-1} = C_5^{(6,q)}$, $(C_2^{(6,q)})^{-1} = C_4^{(6,q)}$ and $(C_3^{(6,q)})^{-1} = C_3^{(6,q)}$, with straightforward but tedious calculations, we have the following Tables 2 and 3. Since $d_{D_i,D_j}(w) = d_{D_j,D_i}(N-w)$ and $C_{s_i,s_j}(w) = C_{s_j,s_i}(N-w)$, we consider only the cases with $i > j$. In these two tables, $C_i^{(6,q)}$ is abbreviated to $C_i$ and $s_i$ is the characteristic sequence of $D_i$. The entry $[a,b]$ in the first column means $i = a$ and $j = b$, respectively. For the other columns, we mean that $d_{D_i,D_j}(w)$ is the top number and $C_{s_i,s_j}(w)$ is the bottom number when $w$ belongs to the set at the corresponding row.

**Table 2:** The cross-correlation functions of Hall sequences.

| $d_{D_i,D_j}(w)$ | $\frac{q-1}{6}$ | $\frac{9q+6x-87}{36}$ | $\frac{9q+6x-15}{36}$ | $\frac{9q+6x+57}{36}$ | $\frac{9q-12x+21}{36}$ | $\frac{9q-12x-51}{36}$ |
|---|---|---|---|---|---|---|
| $[1,0]$ | $\{0\}$ | $C_0$ | $C_5 \cup C_4$ | $C_3$ | $C_2$ | $C_1$ |
| $[2,0]$ | $\{0\}$ | $C_3$ | $C_5 \cup C_2$ | $C_1$ | $C_4$ | $C_0$ |
| $[2,1]$ | $\{0\}$ | $C_1$ | $C_0 \cup C_5$ | $C_4$ | $C_3$ | $C_2$ |
| $[3,1]$ | $\{0\}$ | $C_4$ | $C_0 \cup C_3$ | $C_2$ | $C_5$ | $C_1$ |
| $[3,2]$ | $\{0\}$ | $C_2$ | $C_0 \cup C_1$ | $C_5$ | $C_4$ | $C_3$ |
| $[4,0]$ | $\{0\}$ | $C_4$ | $C_0 \cup C_3$ | $C_2$ | $C_5$ | $C_1$ |
| $[4,2]$ | $\{0\}$ | $C_5$ | $C_4 \cup C_1$ | $C_3$ | $C_0$ | $C_2$ |
| $[4,3]$ | $\{0\}$ | $C_3$ | $C_2 \cup C_1$ | $C_0$ | $C_5$ | $C_4$ |
| $[5,0]$ | $\{0\}$ | $C_2$ | $C_0 \cup C_1$ | $C_5$ | $C_4$ | $C_3$ |
| $[5,1]$ | $\{0\}$ | $C_5$ | $C_4 \cup C_1$ | $C_3$ | $C_0$ | $C_2$ |
| $[5,3]$ | $\{0\}$ | $C_0$ | $C_5 \cup C_2$ | $C_4$ | $C_1$ | $C_3$ |
| $[5,4]$ | $\{0\}$ | $C_4$ | $C_3 \cup C_2$ | $C_1$ | $C_0$ | $C_5$ |
| $C_{s_i,s_j}(w)$ | $\frac{4-q}{3}$ | $\frac{-4x-11}{3}$ | $\frac{2x+1}{3}$ | $\frac{13-4x}{3}$ | $\frac{2x-23}{3}$ | $\frac{2x+25}{3}$ |

**Table 3:** The cross-correlation functions of Hall sequences, continued.

| $d_{D_i,D_j}(w)$ | $\frac{q-1}{3}$ | $\frac{9q-6x+33}{36}$ | $\frac{9q-6x-75}{36}$ | $\frac{9q+12x+33}{36}$ | $\frac{9q-6x-39}{36}$ | $\frac{9q-6x-3}{36}$ | $\frac{9q+12x-75}{36}$ |
|---|---|---|---|---|---|---|---|
| $[3,0]$ | $\{0\}$ | $C_0$ | $C_5$ | $C_4$ | $C_3$ | $C_2$ | $C_1$ |
| $[4,1]$ | $\{0\}$ | $C_1$ | $C_0$ | $C_5$ | $C_4$ | $C_3$ | $C_2$ |
| $[5,2]$ | $\{0\}$ | $C_2$ | $C_1$ | $C_0$ | $C_5$ | $C_4$ | $C_3$ |
| $C_{s_i,s_j}(w)$ | $\frac{q+2}{3}$ | $\frac{17-2x}{3}$ | $\frac{-2x-19}{3}$ | $\frac{4x+17}{3}$ | $\frac{-2x-7}{3}$ | $\frac{5-2x}{3}$ | $\frac{4x-19}{3}$ |

By simple calculations, we find that only if $x = -2$, some of the six values of $C_{s_i,s_j}(w)$ in Table 2 may coincide, and when $x = -2$ or $4$, $C_{s_i,s_j}(w)$ in Table 3 takes six different values, which proves the theorem. □

**Example 3.6** *When $q = 31$, $(x = -2)$, the cyclotomic classes of order $6$ is $C_0 = \{1, 2, 4, 8, 16\}$, $C_1 = \{3, 6, 12, 17, 24\}$, $C_2 = \{5, 9, 10, 18, 20\}$, $C_3 = \{15, 23, 27, 29, 30\}$, $C_4 = \{7, 14, 19, 25, 28\}$ and $C_5 = \{11, 13, 21, 22, 26\}$. It is easy to get $s_i$, $i = 0, 1, 2, 3, 4, 5$, as follows.*

$$s_0 = (0,1,1,1,1,0,1,0,1,0,0,0,1,0,0,1,1,1,0,0,0,0,0,1,1,0,0,1,0,1,1),$$
$$s_1 = (0,0,0,1,0,1,1,1,0,1,1,0,1,0,1,0,0,1,1,1,1,0,0,0,1,1,0,0,1,0,0),$$
$$s_2 = (0,0,0,0,0,1,0,0,0,1,1,1,0,1,0,1,0,0,1,0,1,1,1,1,0,0,1,1,0,1,1),$$
$$s_3 = (0,1,1,0,1,0,0,1,1,0,0,0,0,0,1,1,1,0,0,1,0,0,0,1,0,1,0,1,1,1,1),$$
$$s_4 = (0,1,1,0,1,0,0,1,1,0,0,1,0,1,1,0,1,0,0,1,0,1,1,0,0,1,1,0,1,0,0),$$
$$s_5 = (0,1,1,0,1,1,0,0,1,1,1,1,0,1,0,0,1,0,1,0,1,1,1,0,0,0,1,0,0,0,0).$$

*We have that $C_{s_i,s_j}(w) \in \{-9, -5, -1, 3, 7, 11\}$ for $[i, j] = [3, 0], [4, 1], [5, 2]$ and $C_{s_i,s_j}(w) \in \{-9, -1, 7, 9\}$ for the other cases. Here we also consider only the cases $[i, j]$ with $i > j$.*

# 4 Conclusions

We investigated the cross-correlation properties of Paley sequences, twin prime sequences and Hall sequences, respectively. The cross-correlation functions of these sequences are decided completely and the results show that all of them are with lower cross-correlation values. Furthermore, to get our conclusion, we established the relations between the multiplier group and decimations of the characteristic sequences for an arbitrary difference set, which illustrate the algebraic structure among all the decimations of a sequence.

# Acknowledgement

The authors would like to thank the anonymous reviewers and the Associate Editor for their careful reading as well as the detailed and helpful comments of the manuscript. This work was supported by NSF of China (No. 60473019 and No. 15071005) and by NKBRPC (No. 2004CB318000 and No. 2005CB321902).

# References

[1] L.D. Baumert, *Cyclic Difference Sets*, Lecture Notes in Mathematics 182, Springer-Verlag, Berlin, 1971.

[2] A. Canteaut, P. Charpin, and H. Dobbertin, *Binary m-Sequences with Three-Valued Crosscorrelation: A Proof of Welchs Conjecture*, IEEE Trans. Inform. Theory, vol. 46, NO. 1, Jan. 2000, 4-8.

[3] H. Dobbertin, P. Felke, T. Helleseth, and P. Rosendahl, *Niho type cross correlation functions via Dickson polynomials and Kloosterman sums*, IEEE Trans. Inform. Theory, vol. 52, NO. 2, Feb. 2006, 613C627.

[4] C. Ding, T. Helleseth and H.M. Martinsen, *New families of binary sequences with optimal three-level autocorrelation*, IEEE Trans. Inform. 47(2001), 428-433.

[5] P. Z. Fan and M. Darnell, *Sequence Design for Communications Applications*, New York: Wiley, 1996.

[6] G. Gong and A. M. Youssef, *Cryptographic Properties of the WelchCGong Transformation Sequence Generators*, IEEE Trans Inform. Theory, vol. 48, NO. 11, Nov. 2002, 2837-2846.

[7] M. Hall Jr., *Combinatorial Theory*, Blaisdell Publishing Company, Massachusett, 1967.

[8] D. L. Henk and Q. Xiang, *A Proof of the Welch and Niho Conjectures of Cross-Correlations of Binary m-Sequence*, Finite Fields and Their Applications (7):253-286, (2001).

[9] D. Jungnickel and A. Pott, *Perfect and almost perfect sequences*, Discr. Appl. Math. 95(1999), 331-359.

[10] S . J. Laverty and D. A. Barr, *A Data Aquisition and Analysis System for Voltage Versus Luminous Intensity Transfer Characteristic Measurements in Flat Panel Displays*, IEEE Trans. Instrum. and Measurem., vol. 40, NO. 3, Jun. 1991, 628-632.

[11] J. H. Lee, S. I. Kim, and D. S. Lee, *Fast Cross-Correlation Method For Real Time Detection of Fetal Heart Rate*, Proceedings of the 20th Annual Znternational Conference of the IEEE Engineering in Medicine and Biology Society, Vol. 20, NO. 1,1998, 178-181.

[12] G. J.Ness and T. Helleseth, *Cross Correlation of m-Sequences of Different Lengths*, IEEE Trans Inform. Theory, vol. 52, NO. 4, Apr. 2006, 1637-1648.

[13] J. D. Olsen, R. A. Scholtz, and L. R. Welch, *Bent-function sequences*, IEEE Trans. Inform. Theory, vol. 28. NO. 6, Nov. 1982, 858-864.

[14] A. Pott, *Finite Geometry and Character Theory*, Lecture Notes in Mathematics 1601, Springer-Verlag, Berlin, 1995.

[15] D.V. Sarwate and M.B. Pursley, *Crosscorrelation properties of pseudorandom and related sequences*, Proc. IEEE 68(1980), 593-619.

[16] A. Shaar, C. Woodcock, and P. Davies, *Bounds on the Cross-Correlation Functions of State m-Sequences*, IEEE Trans. Commun., Vol 35, Issue 3, Mar. 1987, 305-312.

[17] T. Storer, *Cyclotomy and Difference Sets*, Markham Publishing Company, Chicago, 1967.

[18] A.Z. Tirkel, *Cross correlation of m-sequences-some unusual coincidences*, IEEE 4th International Symposium on Spread Spectrum Techniques and Applications Proceedings, Vol 3, 22-25 Sept. 1996, 969-973.

[19] A.Z. Tirkel, E.I. Krengel, and T.E. Hall, *Cross-correlation of binary m and GMW arrays*, IEEE Seventh International Symposium on Spread Spectrum Techniques and Applications, volum 2, 2-5 Sept. 2002, 603-607.

[20] N. Y.Yu and G. Gong, *A New Binary Sequence Family With Low Correlation and Large Size*, IEEE Trans Inform. Theory, vol. 52, NO. 4, Apr. 2006, 1624-1636.

[21] X. Zhang, Z. Zhu, and P. Fan, *Intrusion Detection Based on Cross-correlation of System Call Sequences*, Proceedings of the 17th IEEE International Conference on Tools with Artificial Intelligence (ICTAI05).